\documentclass[sigconf]{acmart}
\usepackage{algorithm}
\usepackage{algorithmic}

\AtBeginDocument{%
  \providecommand\BibTeX{{%
    \normalfont B\kern-0.5em{\scshape i\kern-0.25em b}\kern-0.8em\TeX}}}

\newcommand{\edit}[2]{#2}

\newcommand{\sys}{\textsc{ForeCite}}

\copyrightyear{2020}
\acmYear{2020}
\setcopyright{acmlicensed}\acmConference[SIGIR '20]{Proceedings of the 43rd International ACM SIGIR Conference on Research and Development in Information Retrieval}{July 25--30, 2020}{Virtual Event, China}
\acmBooktitle{Proceedings of the 43rd International ACM SIGIR Conference on Research and Development in Information Retrieval (SIGIR '20), July 25--30, 2020, Virtual Event, China}
\acmPrice{15.00}
\acmDOI{10.1145/3397271.3401235}
\acmISBN{978-1-4503-8016-4/20/07}

\begin{document}
\fancyhead{}

\title{High-Precision Extraction of Emerging Concepts from Scientific Literature}

\author{Daniel King$^{1}$, Doug Downey$^{1}$, Daniel S. Weld$^{1,2}$}
\affiliation{\institution{Allen Institute for AI, Seattle, WA$^{1}$}}
\affiliation{\institution{Paul G. Allen School of Computer Science \& Engineering, University of Washington, Seattle, WA$^{2}$}}
\email{daniel,dougd,danw@allenai.org}

\renewcommand{\shortauthors}{Daniel King, Doug Downey, Daniel S. Weld}

\begin{abstract}
Identification of new concepts in scientific literature can help power faceted search, scientific trend analysis, knowledge-base construction, and more, but current methods are lacking. Manual identification cannot keep up with the torrent of new publications, while the precision of existing automatic techniques is too low for many applications.
We present an unsupervised concept extraction method for scientific literature that achieves much higher precision than previous work. Our approach relies on a simple but novel intuition: each scientific concept is likely to be introduced or popularized by a single paper that is disproportionately cited by subsequent papers mentioning the concept. From a corpus of computer science papers on arXiv, we find that our method achieves a Precision@1000 of 99\%, compared to 86\% for prior work, and a substantially better precision-yield trade-off across the top 15,000 extractions. To stimulate research in this area, we release our code and \edit{data.}{data.\footnote{\url{https://github.com/allenai/ForeCite}}}
\end{abstract}



\maketitle

\section{Introduction}
\label{sec:intro}

Scientific Concept Extraction (SCE) aims to automatically extract concepts discussed in the scientific literature. 
For example, given a corpus of information retrieval papers, we would like to extract \emph{Attentive Collaborative Filtering} \citep{Chen2017AttentiveCF} and \emph{Self-Taught Hashing} \citep{Zhang2010SelftaughtHF} as notable scientific concepts. 
Automatic SCE is necessary because, while some scientific terms are already catalogued in existing knowledge bases such as Wikipedia\footnote{\url{www.wikipedia.org}}, the vast majority are not, due to the breadth and rapid progress of science. Three examples from the top concepts output by our method are \emph{ELMo} \citep{peters2018deep}, \emph{gradient penalty} \citep{Gulrajani2017ImprovedTO}, and \emph{asynchronous advantage actor-critic (A3C)} \citep{Mnih2016AsynchronousMF}, which do not have Wikipedia pages, despite being introduced by papers that now have thousands of citations.  Existing methods struggle with the challenge of distinguishing phrases that are simply \emph{associated} with a concept (e.g. \emph{popular conjecture} or \emph{input graph}) from phrases that truly \emph{are} a concept (e.g. \emph{3-SUM Conjecture} or \emph{graph coloring}). Accurate, high-coverage SCE could power many applications, and is a first step toward automatically constructing a comprehensive scientific knowledge base.

We propose a new method for SCE, \sys , based on the intuition that scientific concepts tend to be introduced or popularized by a single paper---one which is disproportionately cited by other papers.  We encode this intuition in a simple unsupervised algorithm that ranks extracted phrases by how well they follow this citation pattern.
In experiments with recent papers from the computer science domain, we find that \sys\ outperforms the CNLC \citep{Haque2011PhrasesAS} and LoOR \citep{Jo2007DetectingRT} methods from previous work, achieving a better Precision-Yield curve across the top 15,000 extractions and much higher precision among the highest-scoring extractions.

In summary, we make the following  contributions:
\begin{enumerate}
    \item We introduce \sys, a simple, unsupervised method for extracting high-precision conceptual phrases from the scientific literature,
    \item We perform quantitative and qualitative evaluation against other SCE methods, showing \sys\ outperforms existing work, improving Precision@1000 from 86\% to 99\%, and
    \item We release the dataset, code, and evaluation annotations used in our experiments.
\end{enumerate}

\section{Task Definition and Background}
\label{sec:background}

SCE is the task of extracting phrases that are scientific concepts from a corpus of academic papers.  Precisely defining {\em concept} is difficult.  In this paper, we define a concept as a phrase that could reasonably have an encyclopedic article (akin to a Wikipedia page) that would be of interest to multiple scientists.  While subjective, this definition was sufficient to achieve the high end of moderate inter-annotator agreement (Section \ref{subsec:interanno}). By our definition, many phrases are obviously correct (e.g. \emph{BERT} \citep{Devlin2019BERTPO} or \emph{deep learning}) or too ambiguous or vague (e.g. \emph{multiple styles} or \emph{deterministic mechanisms}), but there are also many less clear phrases (e.g. \emph{relationship detection} or \emph{shape analysis}). Phrases can be on the fence because they might be too general, or too specific. For example, \emph{sentence} is too general, and \emph{speaker identification performance} too specific. 
See \edit{Appendix \ref{appendix:anno}}{our GitHub repository} for the instructions used by evaluators in our \edit{experiments.}{experiments.\footnote{\url{https://github.com/allenai/ForeCite/EVALUATION_INSTRUCTIONS.md}}}

Two separate areas of work focus on tasks related to, but distinct from, SCE.  Keyphrase extraction (see \citep{Papagiannopoulou2019ARO} for a recent survey) is the task of extracting important topical phrases at the document level.  In contrast, we extract important phrases at the corpus level---and obtain higher precision and more specific phrases compared to reported results on keyphrase techniques.  Our task also differs from {\em topic modeling} performed by e.g. Latent Dirichlet Allocation (LDA) \citep{Blei2003LatentDA}. In LDA, topics are distributions over the full vocabulary, while our extracted concepts are individual phrases.

\subsection{Existing Methods}
While a variety of phrase mining approaches are applicable to SCE \cite{Mihalcea2004TextRankBO, Blei2003LatentDA, Kleinberg2002BurstyAH, Gollapalli2014ExtractingKF, Shubankar2011AFK, Mesbah2018TSENERAI}, the closest prior works to ours are two approaches \citep{Jo2007DetectingRT, Haque2011PhrasesAS} that use the \emph{term citation graph}, an important building block of our approach. The term citation graph is the subgraph of the full citation graph that includes only papers containing a specific term (e.g. the term citation graph for \emph{neural networks} includes exactly the papers that mention \emph{neural networks}, along with all of their citation edges).

Both previous approaches use the intuition that a term citation graph for a concept should be more dense than that of a non-concept.  For example, a paper that mentions the concept \emph{LSTM} is very likely to cite other papers that also mention \emph{LSTM}. This results in a term citation graph for \emph{LSTM} that is more dense than that of a random term. We describe the two previous approaches below. All methods rank a set of phrases by scoring each phrase independently.

The first method we compare with, \emph{LoOR}, uses log-odds \citep{Jo2007DetectingRT}:

\begin{equation}
\nonumber
\begin{split}
    \text{LoOR}(G_t) = \text{log}(P(O(G_t)|\text{concept})) - \text{log}(P(O(G_t)|\text{not concept}))
\end{split}
\label{eq:loor}
\end{equation}

The LoOR score for a term citation graph $G_t$ is the log probability of making the observation $O(G_t)$ given that $G_t$ is a concept, minus the log probability of making the observation $O(G_t)$ given that $G_t$ is not a concept. See \citep{Jo2007DetectingRT} for details.

The second method we compare with, \emph{CNLC}, builds upon LoOR, but is a simpler formula and normalizes for the size of the term citation graph \citep{Haque2011PhrasesAS}:

\begin{equation}
\nonumber
\begin{split}
    & \text{CNLC}(G_t) = \frac{c_t}{n_t} - \frac{c}{N}
\end{split}
\label{eq:cnlc}
\end{equation}

The CNLC score for a term citation graph $G_t$ is the number of citation edges within $G_t$, $c_t$, divided by the number of papers in $G_t$, $n_t$, minus the number of citation edges from $G_t$ to the rest of the corpus, $c$, divided by the number of papers in the full corpus, $N$. See \citep{Haque2011PhrasesAS} for details.

Both methods use different text preprocessing and datasets to generate the set of candidate phrases. Our preprocessing and dataset are detailed in Section \ref{sec:dataset}.

\section{\sys}
\label{sec:method}

\sys\ is based on a different hypothesis about how the term citation graph tends to be structured for a scientific concept.  Specifically, \sys\ assumes that concepts tend to be introduced or popularized by a {\em central} paper; and that other papers discussing the concept cite the central paper. We show that the term citation graph structure resulting from a central paper is a higher-precision signal than the graph density based signal used in prior work. Specifically, the \sys\ concept score:

\begin{equation}
\nonumber
\begin{split}
\sys(G_t) = max_{p \in G_t}\text{log}(f^p_t + 1)\cdot\frac{f^p_t}{f_t}
\end{split}
\label{eq:ours}
\end{equation}

This score is a maximum over papers in a given term citation graph $G_t$, where each paper $p$ is scored based on the number $f^p_t$ of \emph{future} papers that contain the term $t$ {\em and} cite $p$, and $f_t$ the total number of future papers that contain $t$.  The intuition is that more citations within $G_t$ to $p$ is better (the log term), and a higher fraction of papers containing $t$ that cite $p$ is better (the ratio term). There are two additional details to the algorithm: (1) only papers with at least 3 citations are scored, and (2) we sample 500 of the future papers with the phrase \edit{}{in order} to compute the ratio.

\section{Experimental results}
\label{sec:experiment}
We now present experiments measuring precision and yield of \sys\ on SCE, in comparison to methods from prior work.

\subsection{Dataset}
\label{sec:dataset}
Our corpus contains \edit{N/A}{all} arXiv\footnote{\url{www.arxiv.org}} papers \edit{through}{from 1999 to} \edit{August}{September} 2019 in CS.* and stat.ML categories.  We obtain the title, abstract, body text, and citations of the papers from \edit{anonymized}{the Semantic Scholar corpus \citep{Ammar2018ConstructionOT}, which uses ScienceParse\footnote{ \url{https://github.com/allenai/science-parse} and \url{https://github.com/allenai/spv2}}} to perform extraction. \edit{N/A}{Our corpus contains \textasciitilde203,000 papers.} We extract lemmatized noun phrases using spaCy \citep{spacy2} \texttt{en\_core\_web\_md}, and normalize by removing stopwords using NLTK \citep{journals/corr/cs-CL-0205028} English stopwords plus the word "using."

Our candidate phrases include all noun phrases that occur in any title from 1999-2018, resulting in 
\textasciitilde293,000 candidate phrases \edit{N/A}{from \textasciitilde173,000 papers.} We use citation information from papers outside this range, but candidate phrases must occur in at least one title within this range. Candidates are drawn from titles to increase efficiency of our experiments. \edit{Based on informal experiments with candidates from abstracts, we believe results would be similar with other candidate sets}{Restricting candidates to noun phrases that appear in titles does limit the yield of all algorithms, but based on informal experiments with candidates from abstracts, we believe precision would be similar with other candidate sets.\footnote{With candidates from abstracts, we measured p@10000 of 0.92, similar to the 0.93 for title candidates (Table \ref{tab:p_at_x}).}}

\subsection{Human Evaluation}
\label{subsec:interanno}
Evaluating SCE requires an expert and is labor-intensive. Additionally, our methods consider hundreds of thousands of candidate phrases, and we want to focus our evaluation on the fraction of phrases that are highly ranked by the methods, because high precision is required for many applications.  Annotating a static gold set that is independent of the systems' outputs and is still large enough to explore the high-precision regime is intractable.  Given these difficulties, we evaluated highly-ranked phrases from each method's output, and all evaluation was performed by the first author of this paper. We compute inter-annotator agreement with the second author of this paper, first calibrating on a sample of 20 extractions, and then computing agreement on a sample of 100 (both samples are drawn from the annotations used for evaluation in Figure \ref{fig:union_pr_curve}). We achieve raw agreement of 88\% and a Cohen's Kappa of 0.58. This falls into the high end of moderate agreement, which is reasonable given the subjectivity of the task. The disagreements impacted each method fairly equally, and were primarily due to the second evaluator being less generous regarding what qualifies as a concept. Additionally, none of the disagreements fall in the top 5,000 results of our method, reinforcing confidence in the high precision of the top concepts from our method.  We also release all annotations used for evaluation. The disagreements are illustrative of the subjectivity of the task, so we list them here: \emph{road scene}, \emph{downlink}, \emph{stabilizer}, \emph{quasi-polynomial hitting set}, \emph{human demonstration}, \emph{full-diversity}, \emph{local geometric features}, \emph{unit quaternion}, \emph{recurrent model}, \emph{random formula}, \emph{quadratic-time hardness}, and \emph{retail performance}.\footnote{The phrases are difficult to assess in isolation, but we encourage the reader to search for papers mentioning these phrases to see their use in context.} For future work, we would like to further validate our judgements in a user-facing application.

\subsection{Experiments}
As noted above, objective evaluation of SCE is difficult, and previous work has used a variety of different evaluation procedures. \citet{Jo2007DetectingRT} evaluated by inspection, and \citet{Haque2011PhrasesAS} compared concept phrases against librarian-assigned keywords and searches on arXiv. Here, we evaluate the output of each method and measure precision. A measure of recall is not possible without exhaustive gold data, so we focus on the quality of the highly-ranked phrases in terms of Precision@K and Precision-Yield curves.

\begin{table}[t]
    \caption{Precision@K for each method for K $\in$ \{100, 1000, 10000\}, estimated using a sample of size 100.  \sys\ achieves higher precision than the other methods for all K. \edit{N/A}{Results in boldface indicate significantly greater performance ($p<0.05$) than both baselines, computed using the Fisher Exact Test.\protect\footnotemark}}
    \label{tab:p_at_x}
    \footnotesize
    \begin{tabular}{cccl}
      \toprule
       &\small{\sys}&\small{CNLC}&\small{LoOR}\\
      \midrule
      Precision@100&\textbf{1}&0.93&0.91\\
      Precision@1000&\textbf{0.99}&0.86&0.81\\
      Precision@10000&0.93&0.88&0.84\\
    \bottomrule
   \end{tabular}
\end{table}

\begin{figure}[t]
    \centering
    \includegraphics[width=0.47\textwidth]{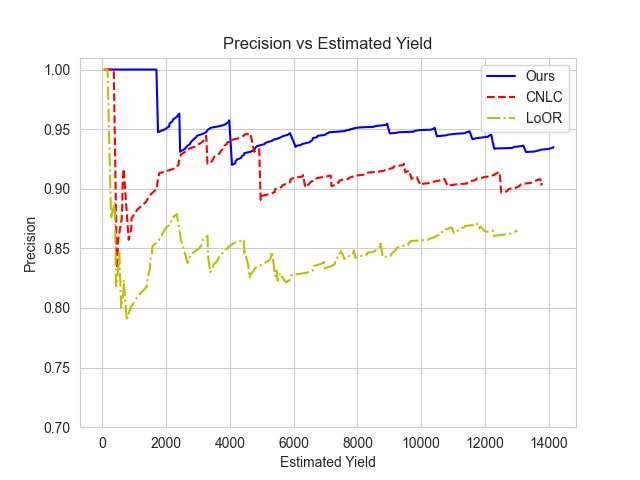}
    \caption{Precision-Yield curves of top 15,000 extractions.  \sys\ outperforms the other methods.}
    \Description{A Precision-Yield curve showing that our method outperforms both CNLC and LoOR across the top 15,000 output concepts from each method}
    \label{fig:union_pr_curve}
\end{figure}

\footnotetext{\url{https://www.socscistatistics.com/tests/fisher/default2.aspx}}

First, we present precision measurements at ranks of 100, 1000, and 10000 in Table \ref{tab:p_at_x}. We evaluated a random sample of size 100 from the top-K ranked phrases of each method. Importantly, even at K of 100, the competing methods do not achieve a precision of 1. Our method does, and maintains precision near 1 out to K of 1,000.

Second, we evaluated a separate random sample of size 300 from the union of the top 15,000 phrases from each method. In Figure \ref{fig:union_pr_curve}, we present a Precision-Yield curve for each method from these annotations. Each point on this curve corresponds to one positive annotation, and has an x value of the estimated true positive yield and a y value of the cumulative precision. Our method outperforms the baselines both at the high-precision end of the curve, and overall, resulting in a 38\% reduction in area over the curve relative to CNLC and a 60\% reduction relative to LoOR.

\section{Discussion}
\label{sec:qual}

\begin{figure}
  \centering
  \begin{tabular}{c c c}
    \includegraphics[width=.46\linewidth]{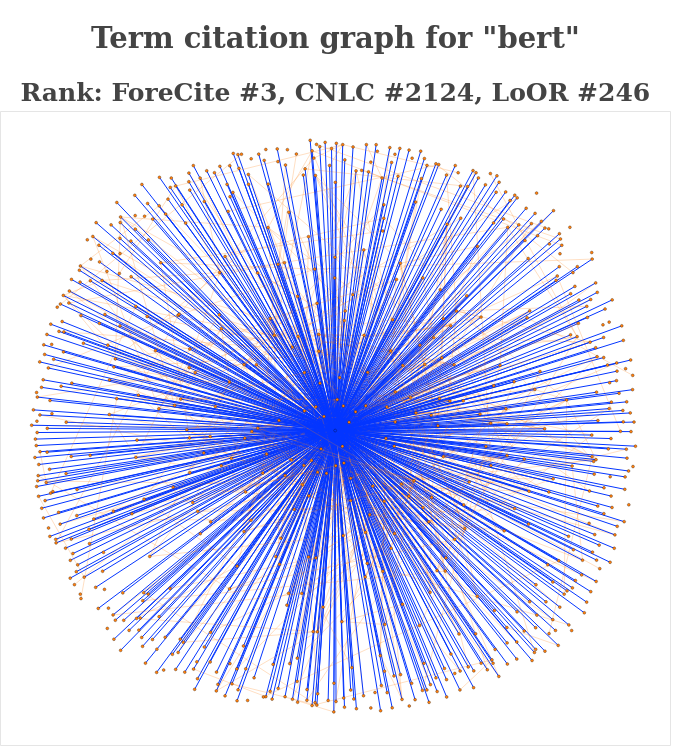} & \includegraphics[width=.46\linewidth]{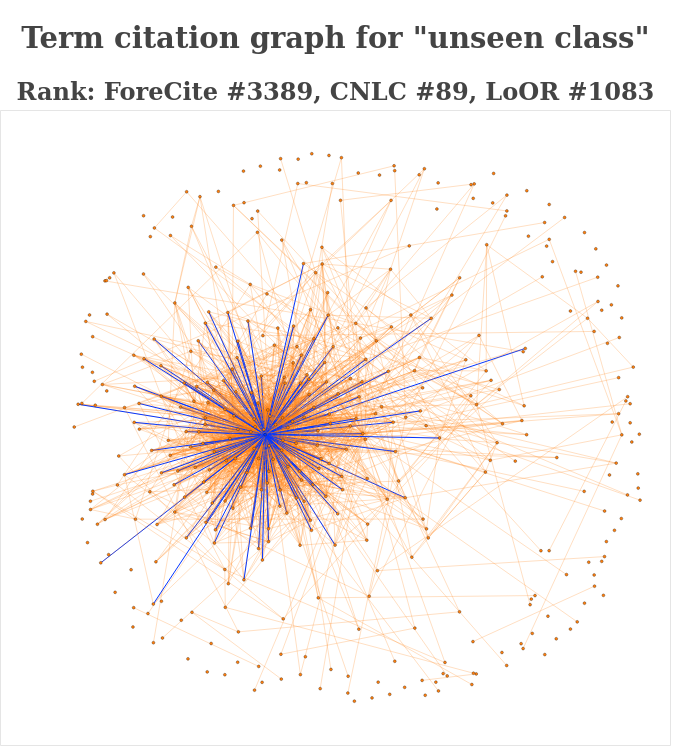} 
  \end{tabular}

  \caption{Illustrative term citation graphs for a concept highly ranked by \sys\ (\emph{BERT}, left), and CNLC (\emph{unseen class}, right) respectively. Blue edges connect to the central paper according to our method, other edges are orange. The \emph{BERT} graph has a much higher proportion of links to the central paper than the \emph{unseen class} graph.}
  \Description{Three plots, each showing the term citation graph for a high-scoring term from one of the methods. The plots show that for our method, a much higher proportion of the edges in the graph are connected to a single central node, in this case, the BERT paper.}
  \label{fig:graph_plots}
\end{figure}

\subsection{Qualitative Method Comparison}

Qualitatively, the different SCE methods favor different types of term citation graphs. We illustrate this difference in Figure \ref{fig:graph_plots}, which plots the term citation graphs for two phrases, one highly ranked by \sys, one highly ranked by CNLC.  \sys\ values a ``spiky'' graph structure with many links to a central paper, while CNLC and LoOR value dense graph structures indicative of a phrase shared amongst a citation community. This difference can be seen in Figure \ref{fig:graph_plots}, where the \emph{BERT} citation graph is dominated by links to the central paper, colored blue (603 out of 678 nodes link to the max degree node), whereas the \emph{unseen class} citation graph has a much higher proportion of links between other papers, colored orange (113 out of 415 nodes link to the max degree node). Additionally, LoOR does not normalize for the size of the term citation graph, so it favors phrases that occur very frequently in the corpus. \edit{N/A}{The above differences between the methods suggest that \sys\ could be a helpful addition to existing tools.}

The intuition behind \sys\ also leads it to produce more specific concepts. For example, the top five phrases ranked by \sys\ in our experiments are \emph{fast gradient sign method}, \emph{DeepWalk}, \emph{BERT}, \emph{node2vec}, and \emph{region proposal network}. By comparison, CNLC's top five are \emph{VQA}, \emph{adversarial example}, \emph{adversarial perturbation}, \emph{ImageNet}, and \emph{person re-identification}, and LoOR's top five are \emph{codeword}, \emph{received signal}, \emph{achievable rate}, \emph{convolutional layer}, and \emph{antenna}. Given our goal of augmenting existing knowledge bases with new, specific, concept pages, we would like to know if we are extracting \emph{emerging} concepts relative to Wikipedia. \edit{N/A}{Wikipedia has specific inclusion criteria\footnote{\url{https://en.wikipedia.org/wiki/Wikipedia:Notability}}, but more recent and specific concepts are less likely to have Wikipedia pages, and these are the type of concepts that \sys\ tends to rank highly.} As an indication of this, we measure how many of the top-20 phrases from each method already have Wikipedia pages, finding that only 30\% of them do for our method, compared to 50\% and 90\% for CNLC and LoOR respectively.

Due to the high precision of \sys\ in the regimes measured in our experiments, our data includes only 18 unique errors for the method. An error analysis revealed that ten of the 18 phrases were too general or vague, whereas the other eight were too specific. Further, five errors were due to mistakes in PDF parsing or noun phrase extraction, rather than \sys 's ranking heuristic.

\subsection{Analysis of Central Papers}
The intuition behind \sys\ is that valid concepts are generally introduced in a central paper, and \sys\ identifies this central paper. As verification of the importance of this intuition, we evaluated our top-100 concepts for whether the central paper does in fact introduce the concept, and found that it does for 95 of them. For example, the concept \emph{fast gradient sign method} is associated with \citet{Goodfellow2015Explaining}, which introduces the \emph{fast gradient sign method} as a way to generate adversarial examples.  The content and citations of the introducing paper are a rich data source for downstream applications such as constructing a knowledge base entry for the concept; exploring this is an item of future work.

Applied to the two most cited papers from SIGIR 2017 (that are on arXiv) according to Microsoft Academic\footnote{\url{www.academic.microsoft.com}}, our method correctly identifies the highest-scoring concept for each paper as \emph{IRGAN} \citep{wang2017irgan} and \emph{Neural Factorization Machine} \citep{he2017neural}.

\section{Conclusion and Future Work}
\label{sec:conclusion}

In this paper, we present a simple, unsupervised method for high-precision concept extraction from the scientific literature. We show that our method outperforms prior work using term citation graphs, particularly in the high-precision regime. In future work, we would like to apply our method to a corpus beyond arXiv computer science, use the output of our method as distant supervision for more powerful textual concept extraction, and use our concept extraction as a starting point for further applications, including semi-automated construction of encyclopedia pages for science.

\begin{acks}
\edit{N/A}{This work was supported in part by NSF Convergence Accelerator award 1936940, ONR grant N00014-18-1-2193, and the University of Washington WRF/Cable Professorship. We thank Iz Beltagy and Oren Etzioni for helpful discussions, Madeleine van Zuylen for assistance with early evaluations, Marti Hearst for comments on a draft, and the rest of the Semantic Scholar team for comments on drafts, and assistance with the text processing and data collection.}
\end{acks}

\bibliographystyle{ACM-Reference-Format}
\bibliography{bibliography}


\end{document}